\documentclass[showpacs,preprintnumbers,prd,nofootinbib,floats,amssymb,floatfix]{revtex4}
\usepackage{amsmath}
\usepackage{amsfonts}
\usepackage{graphicx}
\usepackage{hyperref}

\usepackage{amsmath}
\usepackage{amstext}
 
\begin{document}

\title { A pure Dirac's method for  Husain-Kuchar theory}  
\author{ Alberto Escalante}  \email{aescalan@sirio.ifuap.buap.mx}
 \affiliation{Instituto de F{\'i}sica Luis Rivera Terrazas, Benem\'erita Universidad Aut\'onoma de Puebla, (IFUAP).
   Apartado postal      J-48 72570 Puebla. Pue., M\'exico, } 
   \author{J. Berra}
 \affiliation{ Facultad de Ciencias F\'{\i}sico Matem\'{a}ticas, Benem\'erita  Universidad Au\-t\'o\-no\-ma de Puebla,
 Apartado postal 1152, 72001 Puebla, Pue., M\'exico.}

\begin{abstract}

A pure Dirac's canonical analysis, defined in the full phase space for the Husain-Kuchar model is discussed in detail. This approach allows us to determine the extended action, the extended Hamiltonian, the complete constraint algebra and the gauge transformations for all variables that occur in the action principle.  The complete set of constraints defined on the full phase space  allow us to calculate the Dirac algebra structure of the theory and a local weighted measure for the path integral quantization method.  Finally, we discuss  briefly the necessary mathematical structure to perform the canonical quantization program within the framework of the loop quantum gravity approach.
\end{abstract} 
\date{\today}
\preprint{}
\maketitle

\section{ INTRODUCTION}
The construction of a consistent quantum theory of gravity is a  difficult task. It has been a long way since the  first quantized geometric models,  including  topological field theories, lower dimensional gravity and  minusuperspace homogeneus cosmological models  \cite{Witten, Ryan}.  Loop quantum gravity (LQG) has emerged in recent years as one of the most important candidates  for describing the unification between   gravity and quantum mechanics. The theory has a mathematical rigorous basis for its quantum kinematics \cite{kinematics} given by  the  measure defined in the configuration space \cite{uniqueness}. It also has achieved several promising physical results, at the theoretical level it provides a detailed microscopic picture of black hole entropy and the big bang scenario  in the context of homogeneous quantum cosmologies. Nevertheless, the problem of the dynamics of the full theory has remained unsettled. In this respect, there are    open issues as  for instance how general relativity (GR) arises from LQG as a semiclassical limit of the quantum theory, and the fact that the algebra of quantum constraints, though free of anomalies, does not correspond to the algebra of classical constraints completely \cite{Perez}. Furthermore,  other  problems are centered  about   the Hamiltonian constraint \cite{Hamconst} which generates the dynamics of the system,  unlike the gauge constraints associated with spatial diffeomorphisms and the internal gauge group remaining  under control. Thus,  the problem with the Hamiltonian constraint   within the quantization procedure   remains controversial. In this manner, to understand and solve  the problems found in the quantization of  gravity, has been necessary to study models with a close relation with GR,  such is the case of gravity in three dimensions,  topological field theories,  $BF$ field theories, or an  interesting model  the so-called Husain-Kuchar (HK) model \cite{Husain}. HK  theory is a background independent theory, it share relevant   symmetries  used in the quantum regime of GR as  for instance  the  diffeomorphisms covariance, however,   it lacks of  the hamiltonian constrain. Hence, HK model  is  a four-dimensional background-independent system with local degrees of freedom,  describing  equivalence classes of metrics in the spatial slices of a 3+1 foliation of space-time, although without time evolution \cite{Barbero}. In other words, it is like trying to obtain information from the whole space-time manifold by studying the embedded hypersurfaces. Even so, it is believed that a quantization of this kind of covariant field theories would shed light towards the quantization of  GR. With  respect to the hamiltonian analysis, HK theory is   very close to 3+1 GR expressed in terms of Ashtekar variables \cite{Ashtekar}. In fact, the solutions to  Einstein's equations with ${\rm SO}(3)$ as the internal group,  can be seen as a subset of the solutions to  HK theory.  \\
In the present letter,   we perform a \textit{pure Dirac's analysis}  to HK model. By this we mean that we will  consider all the variables that occur in the Lagrangian density as dynamical variables and not only those ones that occur in the action with  temporal derivatives \cite{Escalante, Jasel}. One could naively realize that a pure  Dirac's analysis is not mandatory, however it is not true at all. The approach developed in this work,  is  quite different to the standard Dirac's analysis; this means that in agreement with the background independence structure that presents the theory under study, we will develop
the Hamiltonian framework by considering all the  fields defining our  theory  as dynamical ones; this fact will allow us to  find the complete structure of the constraints, the equations of motion,
gauge transformations, the extended action as well as the extended Hamiltonian. Generally,  in theories just like GR  we are  able to realize
that developing the Hamiltonian approach on a smaller phase space context, the structure obtained
for the constraints is not right. In fact, we observe in \cite{Peldan} that the Hamiltonian constraint for
Palatini theory does not has the required structure to form a closed algebra with all constraints;
this problem emerges because  by working on a smaller phase space context we do not  have  control on
the constraints,  in order to obtain the correct structure of them they have usually   fixed by hand. Moreover,   in  three dimensional tetrad gravity, in despite of the existence of several  articles performing the hamiltonian analysis on the reduced phase space, in some papers
it is written that the gauge symmetry is Poincare symmetry \cite{Witten, 1}, in others that is Lorentz symmetry plus diffeomorphisms \cite{Carlip}, or that  there exist various ways to define the constraints leading to different gauge transformations. We think that   a pure Dirac's formalism is the best  tool  for solving these problems and for [HK] theory we have performed an a complete analysis.   Nevertheless,  by working with the full phase space  it is possible to find the  full set of constraints  defined on the full phase space,  allowing us to obtain for instance,  a closed algebra among the constraints  just as  is required   \cite{Escalante}. For these reasons,  we believed that a complete Dirac's approach  applied   to  HK theory would be useful for understanding  the problems  found in  GR. Finally, it is worth mentioning that once the full set of constraints is calculated, this procedure could shed light on the search of observables in the context of covariant field theories, specifically in the case of strong-Dirac observables, which must be defined on the complete phase space and not on the reduced one. Finally, our approach will allow us determine  the  measure in the path integral quantization approach, all those  ideas will be clarified along the paper\\
\noindent        
The paper is organized as follows: In Section II, we perform the pure Dirac's method for the HK model, and we find the extended Hamiltonian, the extended action, the complete constraint algebra, the full phase space gauge transformations, the Dirac bracket structure and the path integral measure. In Section III  we discuss on remarks and conclusions. 

\section{A pure Dirac's method for the Husain-Kuchar model}

In this section, we carry out a pure Dirac's analysis. As was  commented above, the approach developed in this work consists in consider  that   all the variables that occurs  in the action will be treated as  dynamical ones,  of course, our approach is fully  in agreement with the background independence  of the theory because all the fields that  define our theory  are fully dynamical, none are fixed.\\
We start with the following  action principle \cite{Husain} 
\begin{equation}\label{action}
S\left[ e,A\right]=\int_{\mathcal{M}}\epsilon_{IJK}\;e^{I}\wedge e^{J}\wedge F^{K}\left[ A\right],  
\end{equation}

\noindent where $F^{I}\left[ A\right]=dA^{I}+\epsilon^{I}_{\;JK}A^{J}\wedge A^{K}$ is the curvature of the $su(2)$-valued connection 1-form $A^{I}=A_{\alpha}^{I}dx^{\alpha}$, which defines a covariant derivative acting on the gauge group $D_a \lambda^I_b= \partial_a \lambda^I_b + \epsilon{^I}_{JK} A^J e^K_b$, and $e^{I}=e_{\mu}^{I}dx^{\mu}$ is a $su(2)$-valued 1-form field. Here $\mathcal{M}$, is a four dimensional manifold $\mathcal{M}=\mathbb{R}\times \Sigma$, with $\Sigma$ (which we take to be compact and without boundary) corresponding to Cauchy's surfaces and $\mathbb{R}$ representing an evolution parameter. Capital letters indices represent the internal space, greek indices are spacetime indices running over 0,1,2,3, and $x^{\mu}$ are the coordinates that label the points on the manifold $\mathcal{M}$. The internal group manifold carries an $su(2)$-invariant metric, $\delta_{IJ}$, and a volume element given by the Levi-Civita tensor, $\epsilon_{IJK}$. It is important to remark that  there exist alternative proposals  to  action  (\ref{action}) \cite{Barbero, Merced}, which express  to  HK model as a constrained BF-like  theory. The equations of motion that arise from the variation of the action are given by
\begin{eqnarray}
\epsilon_{IJK}e^{J}_{[a}F_{bc]}^{K}&=&0, \nonumber\\
\epsilon_{IJK}e^{J}_{[a}D_{b}e^{K}_{c]}&=&0.
\end{eqnarray}

\noindent In this sense, both $e$ and $A$ are considered as dynamical variables and we will take this fact into account below.
\noindent By performing the $3+1$ decomposition, we can write the action as
\begin{equation}
S\left[e,A \right]=\int_{\mathit{R}\times\Sigma}\eta^{ijk}\epsilon_{IJK}\left[e_{0}^{I}e_{i}^{J}F_{jk}^{K}+e_{i}^{I}e_{j}^{J}F_{0k}^{K} \right]  
\end{equation}

\noindent where $\eta^{ijk}=\epsilon^{0ijk}$. From this action, we identify the Lagrangian density
\begin{equation}
\mathcal{L}=\epsilon_{IJK}\left[e_{0}^{I}e_{i}^{J}F_{jk}^{K}+e_{i}^{I}e_{j}^{J}F_{0k}^{K} \right].  
\end{equation}

\noindent By determining the set of dynamical variables, the application of the pure Dirac's method calls for the definition of the momenta $\left(p^{\alpha}_{A}, \pi^{\alpha}_{A} \right) $, 
\begin{equation}
p^{\alpha}_{A}=\frac{\delta\mathcal{L}}{\delta \dot{e}^{A}_{\alpha}}, \quad\quad \pi^{\alpha}_{A}=\frac{\delta\mathcal{L}}{\delta\dot{A}^{A}_{\alpha}},
\end{equation}

\noindent canonically conjugate to $\left(e^{A}_{\alpha},A^{A}_{\alpha} \right) $. The matrix elements of the Hessian,
\begin{equation}
\frac{\partial^2{\mathcal{L}} }{\partial (\partial_\mu A_{\alpha  }^{A} ) \partial(\partial_\mu A_{\beta }^{B} ) }, \quad \frac{\partial^2{\mathcal{L}} }{\partial (\partial_\mu e^{A}_{\alpha} ) \partial(\partial_\mu A_{\beta }^{B} ) }, \quad \frac{\partial^2{\mathcal{L}} }{\partial (\partial_\mu e^{A}_{\alpha} ) \partial(\partial_\mu e^{B}{\beta } ) },
\end{equation}

\noindent vanish, which means that the rank of the Hessian is equal to zero, so that,  24 primary constraints are expected. From the definition of the momenta, it is possible to identify the following 24 primary constraints:

\begin{eqnarray}
\phi^{0}_{A}&:& p^{0}_{A} \approx 0, \nonumber \\
\phi^{a}_{A}&:& p^{a}_{A} \approx 0, \nonumber \\
\psi^{0}_{A}&:& \pi^{0}_{a} \approx 0, \nonumber \\
\psi^{a}_{A}&:& \pi^{a}_{A}-\eta^{abc}\epsilon_{ABC}e^{B}_{b}e^{C}_{c} \approx 0. 
\end{eqnarray}

\noindent By neglecting the terms on the frontier, the canonical Hamiltonian for the HK model is expressed as
\begin{equation}
H_{\rm c}=-\int_{\Sigma}dx^{3}\left[A_{0}^{I}D_{i}\pi^{i}_{I}+\eta^{ijk}\epsilon_{IJK}e_{0}^{I}e_{i}^{J}F_{jk}^{K} \right]. 
\end{equation}

\noindent By adding the primary constraints to the canonical Hamiltonian, we obtain the primary Hamiltonian 
\begin{equation}
H_{\rm P}=H_{\rm c}+\int_{\Sigma}dx^{3}\left[ \lambda_{0}^{I}\phi^{0}_{I}+\lambda^{I}_{i}\phi^{i}_{I}+\gamma^{I}_{0}\psi^{0}_{I}+\gamma^{I}_{i}\psi^{i}_{I}\right],   
\end{equation}

\noindent where $\lambda^{I}_{0}$, $\lambda^{I}_{i}$, $\gamma^{I}_{0}$ and $\gamma^{I}_{i}$ are Lagrange multipliers enforcing the constraints. The non-vanishing fundamental Poisson brackets for the theory under study are given by
\begin{eqnarray}
\{e^{A}_{\alpha}(x^{0},x),  p^{\mu}_{I}(y^{0},y)  \} & =& \delta^\mu_\alpha \delta^{A}_{I}  \delta^3(x,y), \nonumber \\
\{ A^{A}_{\alpha}(x^{0},x), \pi^{\mu}_{I}(y^{0},y) \} &=&  \delta^\mu_\alpha \delta^{A}_{I} \delta^3(x,y).
\end{eqnarray}

\noindent Now, we need to identify if the theory has secondary constraints. For this aim, we compute the $24\times 24$ matrix whose entries are the Poisson brackets among the primary constraints

\begin{eqnarray}\label{primary}
\{ \phi^{0}_{A}(x),\phi^{0}_{I}(y) \}&=&0,   \qquad   \{ \phi^{a}_{A}(x),\phi^{0}_{I}(y) \} = 0  \nonumber \\
\{ \phi^{0}_{A}(x),\phi^{i}_{I}(y) \} &=& 0, \qquad  \{ \phi^{a}_{A}(x),\phi^{i}_{I}(y) \} = 0, \nonumber \\
\{ \phi^{0}_{A}(x),\psi^{0}_{I}(y) \} &=& 0, \qquad  \{ \phi^{a}_{A}(x),\psi^{0}_{I}(y) \} = 0, \nonumber \\
\{ \phi^{0}(x)_{A},\psi^{i}_{I}(y) \} &=& 0, \qquad  \{ \phi^{a}_{A}(x),\psi^{i}_{I}(y) \} = 2\eta^{aik}\epsilon_{AIK}e^{K}_{k}(y)\delta(x,y), \nonumber \\ 
\{ \psi^{0}_{A}(x),\phi^{0}_{I}(y) \}&=&0,   \qquad   \{ \psi^{a}_{A}(x),\phi^{0}_{I}(y) \} = 0  \nonumber \\
\{ \psi^{0}_{A}(x),\phi^{i}_{I}(y) \} &=& 0, \qquad  \{ \psi^{a}_{A}(x),\phi^{i}_{I}(y) \} = -2\eta^{aic}\epsilon_{AIC}e^{C}_{c}(x)\delta(x,y), \nonumber \\
\{ \psi^{0}_{A}(x),\psi^{0}_{I}(y) \} &=& 0, \qquad  \{ \psi^{a}_{A}(x),\psi^{0}_{I}(y) \} = 0, \nonumber \\
\{ \psi^{0}_{A}(x),\psi^{i}_{I}(y) \} &=& 0, \qquad  \{ \psi^{a}_{A}(x),\psi^{i}_{I}(y) \} =0. \nonumber \\
\end{eqnarray}

\noindent This matrix has rank=18 and 6 linearly independent null-vectors, which implies that there are 6 secondary constraints. By requiring consistency of the temporal evolution of the constraints and using the 6 null vectors, the following 6 secondary constraints arise

\begin{eqnarray}
\dot{\phi}^{0}_{A}&=&\{\phi^0_{A}, H_{P} \}\approx 0 \quad \Rightarrow \quad F_{A}:= \eta^{abc}\epsilon_{ABC}e^{B}_{a}F_{bc}^{C} \approx 0,  \nonumber \\
\dot{\psi}^{0}_{A}&=&\{\psi^{0}_{A}, H_P \}\approx 0 \quad \Rightarrow \quad G_{A}:= D_{a}\pi^{a}_{A} \approx 0,
\end{eqnarray}

\noindent and the following Lagrange multipliers are fixed

\begin{eqnarray}\label{multipliers}
\dot{\phi}^{a}_{A}&=&\{\phi^{a}_{A}, H_P \} \approx 0 \quad \Rightarrow \quad \eta^{aik}\epsilon_{AIK}\left(e_{0}^{I}F_{ik}^{K}-\gamma_{i}^{I}e^{K}_{k}  \right)=0 ,  \nonumber \\
\dot{\psi}^{a}_{A}&=&\{\psi^{a}_{A}, H_P \}\approx 0 \quad \Rightarrow \quad 2\eta^{aij}\epsilon_{AIJ}\left[ D_{k}\left( e_{0}^{I}e_{i}^{I}\right)-\lambda^{I}_{i}e^{J}_{i} \right]-\epsilon_{AI}^{\;\;\;\;\;J}A_{0}^{I}\pi^{a}_{J}=0 .
\end{eqnarray}

\noindent This theory does not have terciary constraints. By following  the study, we determine which ones  constraints are first class and which are second class. To accomplish such a task we calculate the Poisson brackets between the primary and secondary constraints. To complete the constraint matrix, we add to the algebra shown in Eq. (\ref{primary})  the following expressions 

\begin{eqnarray}\label{secondary}
\{ \phi^{0}_{A}(x),G_{I}(y) \}&=&0,   \qquad   \{ \phi^{a}_{A}(x),G_{I}(y) \} = 0  \nonumber \\
\{ \phi^{0}_{A}(x),F_{I}(y) \} &=& 0, \qquad  \{ \phi^{a}_{A}(x),F_{I}(y) \} = 0, \nonumber \\
\{ \psi^{0}_{A}(x),G_{I}(y) \}&=&0,   \qquad   \{ \psi^{a}_{A}(x),G_{I}(y) \} = \epsilon_{AI}^{\;\;\;\;K}\pi^{a}_{K}  \nonumber \\
\{ \psi^{0}_{A}(x),F_{I}(y) \} &=& 0, \qquad  \{ \psi^{a}_{A}(x),\phi^{i}_{I}(y) \} =-2\eta^{aij}\left[  \epsilon_{AIJ}e_{i}^{J}(y)\partial_{j}(y)+\epsilon_{AKM}\epsilon_{IJ}^{\;\;\;\; K}e_{i}^{J}A^{M}_{j}(y)\right]\delta(x,y), \nonumber \\
\{ F_{A}(x),F_{I}(y) \}&=&0,   \qquad   \{ G_{A}(x),F_{I}(y) \} = \epsilon_{AI}^{\;\;\;\;C}F_{C}=0  \nonumber \\
\{ G_{A}(x),G_{I}(y) \}&=&\epsilon_{AI}^{\;\;\;\;C}G_{C}=0.
\end{eqnarray}
\noindent The matrix formed by the Poisson brackets between all the constraints exhibited in Eqs. (\ref{primary}) and (\ref{secondary}) has rank=18 and 12 null-vectors \footnote{The null vector space generated by the complete constraint hypersurface is a subset of a quotient vector space $\mathbb{R}^{30}/\mathcal{G}$, where $\mathcal{G}$ is the set given by all the primary and secondary constraints.}. The contraction of the null vectors with the matrix formed by the constraints,  one obtains the following  12 first class constraints	

\begin{eqnarray}\label{firstclass}
\Phi^{0}_{A}&:& p^{0}_{A}, \nonumber \\
\Psi^{0}_{A}&:& \pi^{0}_{a}, \nonumber \\
 G_{A}&:& D_{a}\pi^{a}_{A}+\epsilon_{AB}^{\;\;\;\;\;C}e_{a}^{B}p^{a}_{C},\nonumber \\
 F_{A}&:& \epsilon_{ABC}\eta^{abc}e^{B}_{a}F_{bc}^{C}+D_{a}p^{a}_{A}.
\end{eqnarray}

On the other hand, the rank allow us to   find  the following 18 second class constraints
\begin{eqnarray}\label{secondclass}
\chi^{a}_{A}&:& p^{a}_{A}\approx 0, \nonumber \\
\xi^{a}_{A}&:& \pi^{a}_{A}-\eta^{abc}\epsilon_{ABC}e^{B}_{b}e^{C}_{c} \approx 0.
\end{eqnarray}
 
\noindent The correct identification of the constraints
is a very important step because they are used to carry out the counting of the physical degrees of freedom
and to identify the gauge transformations if there exist  first class constraints. On the other hand, the constraints are the guideline to make the best progress for the quantization of the theory. Hence,  the counting of degrees of freedom is carry out  follows: there are 48 canonical variables, 12 independent first class constraints and 18 independent second class constraints, which leads to determine, that theory under study   has 3 degrees of freedom per space-time point. Of course, by considering the second class constraints Eq. (\ref{secondclass}) as strong equations, the above relations are reduced to the usual constraints obtained in \cite{Husain}, so  this analysis extends and completes those  results found  in the literature.

\noindent By calculating the algebra among the constraints, we find that

\begin{eqnarray}\label{algebra}
\{ \Phi^{0}_{A}(x),\Phi^{0}_{I}(y) \}&=&0,   \qquad   \{ \chi^{a}_{A}(x),\chi^{i}_{I}(y) \} = 0, \nonumber \\
\{ \Phi^{0}_{A}(x),\chi^{i}_{I}(y) \} &=& 0, \qquad  \{ \chi^{a}_{A}(x),\Psi^{0}_{I}(y) \} = 0, \nonumber \\
\{ \Phi^{0}_{A}(x),\Psi^{0}_{I}(y) \} &=& 0, \qquad   \{ \chi^{a}_{A}(x),\xi^{i}_{I}(y) \} =2\eta^{aik}\epsilon_{AIK}e^{K}_{k}(y)\delta(x,y), \nonumber \\ 
\{ \Phi^{0}_{A}(x),\xi^{i}_{I}(y) \} &=& 0, \qquad  \{ \Psi^{0}_{A}(x),\Psi^{0}_{I}(y) \} = 0, \nonumber \\
\{ \Psi^{0}_{A}(x),\xi^{i}_{I}(y) \} &=& 0, \qquad \{ \xi^{a}_{A}(x),G_{I}(y) \} = \epsilon_{AI}^{\;\;\;\;K}\xi^{a}_{K}\approx 0, \nonumber \\
\{ \Phi^{0}_{A}(x),F_{I}(y) \} &=& 0, \qquad  \{ \chi^{a}_{A}(x),G_{I}(y) \}=\epsilon_{AI}^{\;\;\;\;\;C}\chi^{a}_{C}\approx 0,  \nonumber \\
\{ \Psi^{0}_{A}(x),G_{I}(y) \}&=&0, \qquad \{ \chi^{a}_{A}(x),F_{I}(y) \}=F_{A}e^{a}_{I}+\chi^{i}_{A}D_{i}e^{a}_{I}\approx 0,\nonumber \\
  \{ \Phi^{0}_{A}(x),G_{I}(y) \}&=&0, \qquad  \{ \xi^{a}_{A}(x),F_{I}(y) \} = G_{I}e^{a}_{A}-\epsilon_{IB}^{\;\;\;\;C}e^{a}_{A}e^{B}_{b}\chi^{b}_{C} \approx 0,  \nonumber \\
 \{ \Psi^{0}_{A}(x),F_{I}(y) \}&=&0, \qquad \{ G_{A}(x),F_{I}(y) \}=\epsilon_{AI}^{\;\;\;\;C}F_{C} \approx 0, \nonumber \\
\{ F_{A}(x),F_{I}(y) \}&=&0, \qquad \{ G_{A}(x),G_{I}(y) \}=\epsilon_{AI}^{\;\;\;\;C}G_{C} \approx 0,
\end{eqnarray}

\noindent from where we  appreciate that the constraints form a set of first and second class constraints, as is  expected.  The obtention of the  constraints  defined on the full phase space, will  allow us to find the extended action. By employing the first class constraints  (\ref{firstclass}), the second class constraints  (\ref{secondclass})  and the Lagrange multipliers  (\ref{multipliers}), we find that the extended action takes the form
\begin{eqnarray}
S_{E}[e_{\alpha}^{A},p^{\alpha}_{A},A_{\alpha}^{A},\pi^{\alpha}_{A},\lambda_{0}^{A},\lambda_{a}^{A},\gamma_{0}^{A},\gamma_{a}^{A} ,\lambda^{A},\gamma^{A}] &=& \int [ \dot{e}_{\alpha}^{A}p^{\alpha}_{A}+\dot{A}_{\alpha}^{A}\pi^{\alpha}_{A}-H -\lambda_{0}^{A}\Phi^{0}_{A}-\gamma^{A}_{0}\Psi^{0}_{A}-\lambda^{A}G_{A} \nonumber \\
&&\;\;\;\;\;-\gamma^{A}F_{A}-\lambda^{A}_{a}\chi^{a}_{A}-
\gamma^{A}_{a}\xi^{a}_{A}]d^{4}x, \nonumber 
\end{eqnarray}

\noindent where $H$ is a linear combination of first class constraints, and is given by

\begin{equation}\label{Hamiltonian}
H=A_{0}^{A}\left[ D_{a}\pi^{a}_{A}+\epsilon_{AB}^{\;\;\;\;\;C}e_{a}^{B}p^{a}_{C}\right] +e^{A}_{0}\left[\epsilon_{ABC}\eta^{abc}e^{B}_{a}F_{bc}^{C}+D_{a}p^{a}_{A}\right]  
\end{equation}

\noindent and $\lambda_{0}^{A}$ , $\lambda_{a}^{A}$ , $\gamma_{0}^{A}$ , $\gamma_{a}^{A}$ , $\lambda^{A}$ , $\gamma^{A}$ are the Lagrange multipliers enforcing the first and second class constraints respectively. We are able to observe, by considering the second class constraints as strong equations, that the Hamiltonian  (\ref{Hamiltonian}) is reduced to the usual expression found in the literature \cite{Husain}, which is defined on a reduced phase space context. From the extended action, we identify the extended Hamiltonian  which is given by
\begin{equation}\label{extendedH}
H_{\rm E}=H-\lambda_{0}^{A}\Phi^{0}_{A}-\gamma^{A}_{0}\Psi^{0}_{A}-\lambda^{A}G_{A}-\gamma^{A}F_{A}.
\end{equation}

\noindent By using  our expressions for the complete set of constraints, it is possible to obtain the gauge transformations acting on the full phase space. For this important step, we shall use  Castellani's formalism \cite{Castellani}, which allows us to define the following gauge generator in terms of the first class constraints:
\begin{equation}\label{generator}
G=\int_{\Sigma}\left[ D_{0}\varepsilon_{0}^{A}p^{0}_{A}+D_{0}\zeta_{0}^{A}+\varepsilon^{A}G_{A}+\zeta^{A}F_{A}\right]d^{3}x, 
\end{equation}

\noindent where $\varepsilon_{0}^{A}$, $ \varepsilon^{A}$, $\zeta_{0}^{A}$ and $\zeta^{A}$ are arbitrary continuum real parameters. Thus, we find that the gauge transformations in the phase space are
\begin{eqnarray}
\delta_{0}e_{0}^{I}&=& D_{0}\varepsilon_{0}^{I}, \nonumber \\
\delta_{0}e_{i}^{I}&=&\epsilon^{I}_{\;\;AB}\varepsilon^{A}e^{B}_{i}-D_{i}\zeta^{I}, \nonumber \\
\delta_{0}A_{0}^{I}&=& D_{0}\zeta_{0}^{A}, \nonumber \\
\delta_{0}A_{i}^{I}&=& -D_{i}\varepsilon^{I}, \nonumber \\
\delta_{0}p_{I}^{0}&=& 0, \nonumber \\
\delta_{0}p_{I}^{i}&=& \epsilon_ {IA}^{\;\;\;\;\;C}\left( \varepsilon^{A}p^{i}_{C}+\eta^{aic}\zeta^{A}F_{bc}^{C} \right), \nonumber \\
\delta_{0}\pi_{I}^{0}&=& 0, \nonumber \\
\delta_{0}\pi_{I}^{i}&=& \epsilon_{IA}^{\;\;\;\;\;C}\left(\epsilon^{A}\pi^{i}_{C}-2\eta^{iab}D_{b}\left(\zeta^{A}e_{Ca} \right)-\zeta^{A}p^{i}_{C} \right).
\end{eqnarray}

\noindent In order to recover the diffeomorphisms symmetry, one can redefine the gauge parameters as $-\varepsilon_{0}^{I}=\varepsilon^{I}=-v^{\alpha}A_{\alpha}^{I}$, and $-\zeta^{I}_{0}=\zeta^{I}=-v^{\alpha}e_{\alpha}^{I}$. With this election, the gauge transformations take the form
\begin{eqnarray}\label{gauge}
e_{\alpha}^{'I} &\longrightarrow & e_{\alpha}^{I}+\mathcal{L}_{v}e^{I}_{\alpha}+D_{[\alpha}e^{I}_{\mu]}v^{\mu}, \nonumber \\
A_{\alpha}^{'I} &\longrightarrow & A_{\alpha}^{I}+\mathcal{L}_{v}A^{I}_{\alpha}+v^{\mu}F_{\mu\alpha}^{I},
\end{eqnarray}

\noindent corresponding to diffeomorphism gauge invariance  \cite{Husain, Barbero1}. Some interesting features follow from the gauge orbits. There exist  a vector density , namely  $n^{\alpha}=\hat{n}^{\alpha}/\hat{e}$  \cite{Barbero2} ,  where $\hat{e}$ represents an auxiliary foliation defined by a scalar function $t$ as $\hat{e}=\hat{n}^{\alpha}\partial_{\alpha}t$. The flow lines of $n^{a}$ define a privileged reference frame through $\hat{n}^{\alpha}=\frac{1}{3!}\epsilon^{\alpha\beta\mu\nu}\epsilon_{IJK}e^{I}_{\beta}e^{J}_{\mu}e^{K}_{\nu}$.   Then, by using (\ref{gauge}), it is possible to notice that, in fact, there is no dynamics in the model. The projections of the field equations onto the direction normal to the spatial slices are zero. One can observe,   that the space-time (degenerate) metric  does not change. In other words, every transverse hypersurface has the same intrinsic geometry, which means that every solution to GR is a solution of the HK model.  

\noindent Classically, it is possible to write down explicitly an infinite number of constants of motion, since the Hamiltonian constraint vanishes identically. It is worth mentioning that every invariant  of the three geometry is a constant of motion. This implies that the theory is a Liuoville integrable covariant field theory with local degrees of freedom. On the other hand, the analysis developed above,   shows that the theory is dimensional reduced, so that, the dynamical evolution of the system is determined by the spatial diffeomorphisms. From Eq. (\ref{generator}), the complete spatial diffeomorphism generator acting in the full  phase space is given by
\begin{equation}
H_{i}=  A_{i}^{A}\left( D_{a}\pi^{a}_{A}+\epsilon_{AB}^{\;\;\;\;\;C}e_{a}^{B}p^{a}_{C}\right) +e_{i}^{A}\left( \epsilon_{ABC}\eta^{abc}e^{B}_{a}F_{bc}^{C}+D_{a}p^{a}_{A}\right). 
\end{equation}

\noindent In order to  recovering the usual spatial diffeomorphism generator reported in  \cite{Husain}, it is simple to see that once the second class constraints  (\ref{secondclass}) are solved,  there exists an homomorphism between the Poisson algebra  (\ref{algebra}), and the usual constraint algebra reported in  \cite{Husain}, thorough a dreibein field. \\

\noindent From the constraint analysis, it is important  to determine the Dirac brackets among  our canonical variables.  We need to remember that after Dirac's brackets are constructed, second class constraints can be treated as strong equations \cite{Henneaux}, hence it is an important step in our calculations.  For this aim,  let $D$ the matrix formed by the Poisson brackets between the second class constraints
\begin{equation}\label{secondclassm}
D_{\alpha\beta}=\left(\begin{array}{cc}0 &  \{ \xi^{i}_{I}(x), \chi^{a}_{A}(y) \} \\ \{ \chi^{a}_{A}(x),\xi^{i}_{I}(y) \}  & 0\end{array}\right).
\end{equation}

\noindent The Dirac bracket between two phase space functionals $F$ and $G$ is defined as
\begin{equation}\label{Dbracket}
\{F(x),G(y)\}_{D}\equiv\{F(x),G(y)\}-\int dudv\{A(x),\Psi^{\alpha}(u)\}D^{-1}_{\alpha\beta}\{\Psi^{\beta}(v),G(y)\},
\end{equation}

\noindent where $\{F(x),G(y)\}$ is the usual Poisson bracket between the functionals $F$ and $G$, $\Psi^{\alpha}=(\chi^{a}_{A},\xi^{a}_{A})$ and $D_{\alpha\beta}^{-1}$ are the components of the inverse of the matrix $D$, which take the values, $D^{-1AI}_{ai}=\frac{1}{2e}(\frac{1}{2}e^{A}_{a}e^{I}_{i}-e^{A}_{i}e^{I}_{a})$. By using the definition of the Dirac bracket, the non-trivial canonical relations, \textit{i.e.} those phase space variables that do not Poisson commute with the second class constraints, are given by
\begin{eqnarray}
\{e^{A}_{a},\pi^{i}_{I}\}_{D}&=&0, \qquad \{A^{A}_{a},\pi^{i}_{I}\}_{D}=\delta^{A}_{I}\delta^{i}_{a},  \nonumber \\
\{e^{A}_{a},p^{i}_{I}\}_{D}&=&0, \qquad   \{A^{A}_{a},e^{I}_{i}\}_{D}=D^{-1AI}_{ai}.
\end{eqnarray}

\noindent Briefly, we turn now to the observables issue. An observable in a theory with first class and second class constraints is defined to be a phase space function whose  Dirac's brackets  commutes with all the first class constraints. In this respect, in the HK model there are not   scalar constraint  so that, the observables will be those phase space functionals whose  Dirac's brackets commute with  Gauss and spatial diffeomorphisms constraints. \\
\noindent We will finish this section with some comments about the path integral quantization procedure. An important aspect for  defining a path integral quantum theory,  is the determination of the correct measure. For theories with constraints and some interacting theories, this aim is no-trivial to obtain, and usually is not given by the heuristic Lebesgue measure. The obtention of the mesure, has been an  relevant step in    developing of  spin-foam models, which can be thought of as a path integral version of LQG \cite{Perez, spin foams}. To cut a long story short (see \textit{e.g.} \cite{Henneaux, senjanovic}), the central ingredient for most applications of the path integral is the generating functional, which in our case takes the form
\begin{small}
\begin{equation}\label{path}
Z=\int \mathcal{D}A^{I}_{\mu}\mathcal{D}\pi^{\mu}_{I}\mathcal{D}e^{I}_{\mu}\mathcal{D}p^{\mu}_{I}\delta(\chi^{a}_{A})\delta(\xi^{a}_{A})\sqrt{|D|}\delta(\Phi^{0}_{A})\delta(\Psi^{0}_{A})\delta(G_{A})\delta(F_{A})\sqrt{|E|}\prod_{\alpha}\delta(\zeta_{\alpha}) \exp{i\int dtd^{3}x \pi^{\mu}_{I}\dot{A}^{I}_{\mu}+p^{\mu}_{I}\dot{e}^{I}_{\mu}},
\end{equation}
\end{small}

\noindent here $D$ denotes the determinant of the matrix given in   (\ref{secondclassm}), $\zeta_{\alpha}$ any choice of gauge fixing conditions, $E$ is the square of the determinant of the Poisson brackets between first class constraints and the gauge fixing conditions. In addition $\mathcal{D}q=\prod_{t\in\mathbb{R}}dq(t)$ for all phase space variables. We will drop the exponential of the current in what follows, since it does not affect any of our manipulations, hence we will deal with the partition function $Z=Z[0]$. At the end   we will be  really interested in $Z[j]/Z$, then we can drop overall constant factors from all subsequent formulas. Moreover, we will assume for simplicity that all the gauge fixing conditions are functions independent of the connection $A^{I}_{\mu}$.
\noindent In the equation (\ref{path}), $D$ is the determinant of the Dirac matrix, equation (\ref{secondclassm}), formed by the second class constraints. Therefore $|D|=[\det \{ \chi^{a}_{A}(x),\xi^{i}_{I}(y) \} ]^{2}$, let $C^{ai}_{AI}$ denote this matrix. From the \textit{singular value decomposition theorem} \cite{Lang}, there exist orthogonal matrices $O^{a}_{b}$, $O^{I}_{J}$ such that $O^{b}_{a}O^{I}_{J}e^{J}_{b}$ is diagonal, that is $O^{b}_{a}O^{I}_{J}e^{J}_{b}=\lambda_a\delta^{I}_{a}$. Let $O^{aI}_{bJ}\equiv O^{a}_{b}O^{I}_{J}$, also an orthonormal matrix, we use it to define
\begin{equation}
\widehat{C^{ab}_{IJ}}=O^{aL}_{dI}C^{dc}_{LK}O^{bK}_{cJ}=\sum_{c}\eta^{abc}\epsilon_{IJc}\lambda_{c}.
\end{equation}
\noindent Then $\widehat{C}^{ab}_{IJ}=0$ when $(I=J)$ or $(a=b)$ or $\{a,b\}\neq \{I,J\}$. Reducing by minors we obtain
\begin{equation}
\det{C^{ab}_{IJ}}=2(\lambda_{1}\lambda_{2}\lambda_{3})^{3}=2(\det e^{I}_{a})^{3}=2e^{3},
\end{equation}
\noindent thus, up to an overall factor, $\sqrt{|D|}=e^{3}\equiv V^{3}$. It is so  difficult  to perform  the integrations of   equation  (\ref{path}) in order to  compute transition amplitudes. However if we transform the integral of the HK Lagrangian in terms of the configuration variables, it would be easy to handle. So, by using the reduce phase space technique \cite{Henneaux}, integrating the second class constraints  (they are in fact, primary second class constraints) and taking into account the bijection between $\pi^{i}_{I}$ and $e^{I}_{i}$ when $\det{e^{I}_{i}}\neq 0$. In terms of the extended Hamiltonian (\ref{extendedH}), it is straightforward to obtain 
\begin{small}
\begin{equation}
Z=\int \mathcal{D}A^{I}_{\mu}\mathcal{D}e^{I}_{\mu}V^{6}\sqrt{|E|}\prod_{\alpha}\delta(\zeta_{\alpha})\exp{i\int \epsilon_{IJK}\;e^{I}\wedge e^{J}\wedge F^{K}\left[ A\right]}.
\end{equation}
\end{small}

\noindent It would be desirable to follow the usual way  by employing perturbation theory. However, there is no expansion that disentangle the free theory   from the interaction term. Then, it is not possible to construct a perturbative quantum theory in the usual way. However, Spin foams intend to be a path integral formulation of LQG and theories covariant  under diffeomorphisms \cite{spin foams}, mainly motivated from Feynman's ideas  but appropriately suited to background independence symmetry.  In this paper, we have developed  all the necessary elements to quantize the HK model  within the framework of LQG.  Due to the fact the HK model has not Hamiltonian constraint, HK  theory does not share the usual ambiguities that are present in the quantum scalar constraint of GR. From the covariant quantization perspective, ( the spin foams formalism \cite{spin foams}), it might be possible to study the different simplicial constraints defined by the HK model when it is  written as a constrained BF theory \cite{Barbero, Merced,BFHusain}. Such  analysis should shed light on the relation between canonical and covariant quantizations for covariant field theories, as in the case of loop quantum gravity and spin foam models approaches.

\section{ Conclusions and prospects}

\noindent In this paper, we have consistently performed a pure Dirac's method of  the HK theory. The  analysis    was  carried out in the full phase space, enabled us to identify the extended action, the extended Hamiltonian and the complete set of constraints. Once the constraints were classified as first and second class  by means of  the null vector space defined by the constraints hypersurface, this procedure allowed us to carry out the counting of the degrees of freedom and to calculate the gauge transformations acting in the complete phase space. From the gauge orbits, we realized that the dynamical evolution of the HK model  is given by the full phase space spatial diffeomorphism generator. This means that, classically, it is possible to write down an infinite number of constants of motion, implying that the model is a complete Liouville integrable covariant field theory with local degrees of freedom. One of the main purposes for working on the complete phase space,  lies on the full identification of   the complete set of constraints,  allowing  us to calculate the Dirac algebra structure of the theory  and a local weighted measure for the path integral quantization approach which will be useful in the spin foam formulation.  Finally, we have  introduced  the necessary mathematical structure to complete the canonical quantization program. We can observe that the resulting  physical Hilbert space will be  constituted by the so called spin network states, which are defined on equivalence classes of graphs under diffeomorphisms. \\
Finally,  we would comment that  our approach allow us know  the complete structure of the constraints, thus, we are able to perform   any discrete approach such as spin foam quantization of gravity \cite{spin foams}. It is  pointed out by Gambini et. al. \cite{Pullin},  the inconsistencies  arising  in discretizations methods  are present in spin foam approaches. When one discretizes the action in order to compute the path integral,
one is left with a discrete action whose equations yield an inconsistent theory. In order to avoid these problems, all the variables that occur in the action must be taken into account and no Hamiltonian reduction is allowed, resulting  the presence of  second class constraints just as our approach. In the same way, the discrete path integral measure obtained in the full phase space would be different to that calculated in the  reduced phase-space approach
(which usually  do not have  second class constraints), resulting  new terms originated by parentheses between second class constraints. It is believed  that these 
terms could cure some divergences that arise from the remaining gauge symmetry of the fields. Hence, we believe that our  approach  could shed light in the continuum limit of the discrete covariant methods reported in the literature. All these ideas are in progress and will be reported in forthcoming works.

\noindent \textbf{Acknowledgements}\\[1ex]
This work was supported by CONACyT under grant CB-2010/157641.

\end{document}